\documentclass[12pt]{article}
\renewcommand{\baselinestretch}{1.2}
\usepackage{indentfirst}
\usepackage{amsmath}
\usepackage{amssymb}
\usepackage{amsfonts}
\usepackage{amscd}
\usepackage{amsbsy}
\usepackage{amsthm}
\usepackage{latexsym}
\usepackage{graphicx,color} 
\usepackage[dvipsnames]{xcolor}
\usepackage[colorlinks]{hyperref}
\hypersetup{linkcolor=blue,citecolor=blue,urlcolor=blue}
\def\beq{\begin{eqnarray}}
\def\eeq{\end{eqnarray}}

\def\det{\,\mbox{det}\,}

\DeclareMathOperator{\cx}{\square}

\def\al{\alpha}
\def\be{\beta}

\def\de{\delta}

\def\na{\nabla}

\def\rh{\rho}
\def\si{\sigma}

\def\Ga{\Gamma}
\def\De{\Delta}
\def\La{\Lambda}

\usepackage[symbol]{footmisc} 
\usepackage{float} 
\usepackage{cite}  
\usepackage{titlesec}

\usepackage{ulem} 

\titleformat*{\section}{\large\bfseries}
\titleformat*{\subsection}{\normalsize\bfseries}
\begin{document}

\begin{center}
\renewcommand*{\thefootnote}{\fnsymbol{footnote}} 
{\Large \bf
Semiclassical bounce
\\
with strong minimal assumptions}
\vskip 6mm

{\bf Wagno Cesar e Silva}
\hspace{-1mm}\footnote{E-mail address: \ wagnorion@gmail.com}
\quad
and
\quad
{\bf Ilya L. Shapiro}
\hspace{-1mm}\footnote{E-mail address: \ ilyashapiro2003@ufjf.br}
\vskip 6mm

Departamento de F\'{\i}sica, ICE, Universidade Federal de Juiz de Fora,
\\
Juiz de Fora, 36036-900, Minas Gerais, Brazil
\end{center}
\vskip 2mm
\vskip 2mm


\begin{abstract}

\noindent
We explore the possibility of avoiding cosmological singularity with
a bounce solution in the early Universe. The main finding is that
simple and well-known semiclassical correction, which describes the
mixing of radiation and gravity in the effective action, may provide
an analytic solution with a bounce. The solution requires a positive
beta function for the total radiation term and the contraction of the
Universe at the initial instant. The numerical estimate shows that
the bounce may occur in an acceptable range of energies, but only
under strong assumptions about the particle physics beyond the
Standard Model.
\vskip 3mm

\noindent
\textit{Keywords:} \ Bounce, conformal anomaly, effective action

\end{abstract}

\setcounter{footnote}{0} 
\renewcommand*{\thefootnote}{\arabic{footnote}} 

\section{Introduction}
\label{sec1}

The initial cosmological singularity is considered an important
indication to either modify general relativity (GR) or introduce
exotic forms of ``matter'' with an unusual equation of state
(see, e.g., \cite{Coles}). One may also think about taking into
account the effects of quantum gravity. The last is a direct
consequence of the fact that the Planck density of matter is achieved in
the vicinity of a singularity. In this sense, singularity may be a
kind of window to observe the quantum gravity effects.

The safest way to avoid the singularity is to have a cosmological
solution with a bounce, as pioneered by Tolman in 1931
\cite{Tolman31}.  Starting from the 1970s, there are numerous
bouncing models \cite{Bounce70,StarBo78}, partially related to
the interest in taking quantum effects into account. Since
then, different bouncing cosmological scenarios attracted a lot of
attention (see \cite{Novello2008,Peter2015} for the reviews of
the literature).
In most of the existing models, the bounce is achieved by using
a scalar field with the specially designed potential, or by using
modified gravity actions.
A new recent trend is related to the introduction of the
non-localities into the gravitational action (see, e.g.,
\cite{Biswas2012,Koshelev2013}). The same purpose can be
achieved by taking into account the non-local semiclassical
corrections \cite{AnJu,AnoBo21}.
One of the challenges in building bounce models is to avoid
pathologies related to quantum instabilities \cite{KaroubyBo}.

The conventional assumption is that the consistent theory of quantum
gravity would be as an ultimate solution for the problem of
singularities. The dimensional arguments indicate that the quantum
gravity effects should become relevant at the Planck scale
$M_P \approx 10^{19}\,GeV$.
On the other hand, the effects of quantum matter fields on the
classical gravitational background (semiclassical gravity) may
produce changes in the action of gravity and matter such that the
solution of the effective equations is free of singularity. In this
way, the mentioned window may be closed to the observer from the
later Universe. The purpose of the present paper is to explore this
possibility by constructing the solution with a cosmological bounce,
where the contraction of the Universe goes on until a minimum point,
after which the expansion starts. This minimal point should
correspond to the energy densities far below the Planck density
$M_P^4$, such that the quantum gravity effects and also the possible
higher derivative terms in the gravitational action are
Planck-suppressed and hence irrelevant. Thus, we take
into account only the quantum effects of matter fields.

The first necessary condition to meet a bounce is to have a
decreasing conformal factor of the metric, $a(t)$, at some initial
instant before the bounce. Since the bounce is a form to remove the
singularity, in its vicinity we can assume that the typical distances
are small, the energy density is high, and the quantum effects of
matter fields are relevant. In such a UV regime, the typical energies
are such that all fields are approximately massless. This feature
has the following two consequences:

\textit{i)} One can use a massless approximation for at least
most of the matter fields in the UV limit. For the sake of
simplicity, let us assume that the non-gravitational contents of
the Universe are pure electromagnetic radiation. Later on, we
discuss how other kinds of matter may change the conditions
of the bounce.

\begin{figure}
\begin{center}
\centering
\includegraphics[width=4.5cm,angle=0]{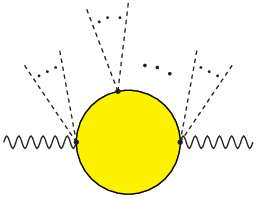}
\caption{\small The loop kernel of matter fields connects to the two
photon lines and an unrestricted number of the dashed lines of the
linearized metric $h_{\mu\nu} =  g_{\mu\nu} - \eta_{\mu\nu}$.}
\end{center}
\label{photon}
\end{figure}

\textit{ii)} Since the matter content can be described by pure
radiation, the
relevant semiclassical diagrams are those with two external lines
of photons and an arbitrary number of linearized gravity tails,
as shown in  Fig.\,\ref{photon}.

In the far UV, when the masses of all quantum fields are small
compared to the energies of the photons (important: not gravitons),
one
can ignore the effect of quantum decoupling, i.e., the Appelquist
and Carazzone theorem \cite{AC}, and take into account the quantum
effects of matter fields by using the minimal subtraction scheme of
renormalization.  In this case, the leading quantum effect is the
conformal anomaly. While the classical radiation decouples from the
dynamical equation for $a(t)$ and affects only the initial condition
(we elaborate on this point below), at the quantum level, the matter
contents of the Universe enter the equation for $a(t)$.

There is extensive literature on the cosmological models based on
conformal anomaly, starting from \cite{fhh} and \cite{star}, where
the anomaly-induced effective action served as an extreme case of the
first inflationary model. However, our present purpose is different,
as we are interested only in the radiation part of the anomaly, which
may become relevant at the lower energies. The explanation is that
the typical energies of the photons in Fig.~\ref{photon} are much
greater than those of gravitons, which define the energy scale of
the vacuum quantum effects. The electromagnetic part of the anomaly
has been used previously in \cite{Dolgov93} to explain the seeds of
magnetic fields in the epoch of forming galaxies. In the present
work, we apply the same approach  to describe the dynamics of the
whole Universe and discuss whether it is sufficient to avoid
singularity in the contracting Universe.

The paper is organized as follows. Section~\ref{sec2} contains a
short survey of the anomaly-induced action in the radiation sector.
In Sec.~\ref{sec3}, we derive the analytic bounce solution in a
theory formed by the Einstein-Hilbert action with the anomalous
contribution mixing the radiation term with gravity. This solution is
supported by the plots obtained using the numerical solution, which
also includes the non-zero cosmological constant case.
Section~\ref{sec4} reports on the numerical estimates for the bounce
and discusses the possibility of overcoming the dramatic physical
inconsistency which we met in the simplest electromagnetic radiation
case. Finally, in Sec.~\ref{sec5}, we draw our conclusions.

We adopt the natural units such that $c=\hbar=1$ and
use the signature $(+,-,-,-)$ for the Minkowski metric $\eta_{\mu\nu}$.

\section{Anomaly-induced action with radiation}
\label{sec2}

It proves useful to present the one-loop beta function for the
square of the gauge coupling $g$ in the unconventional form
 $\be g^4$, where\footnote{
The detailed derivation of this
 expression can be found in many QFT textbooks, e.g., in \cite{OUP}.}
\beq
\be \,=\,-\,
\frac{2}{(4\pi)^2}\,\Big(
\frac{11}{3}\,C_1 - \frac{1}{6}\,N_{cs}
- \frac{4}{3}\,N_f\Big).
\label{ym19}
\eeq
Here $N_{cs}$ and $N_f$ are the numbers of complex scalars and
fermions coupled to the given vector field. $C_1$ is the Casimir
operator of the corresponding gauge group, which is zero in the
Abelian case. The $g^4$ factor was separated from the beta function
for the sake of further convenience. In the non-Abelian theory, $C_1$
is positive, and this opens the possibility of the asymptotic freedom
in the theory \cite{GrossWilczek,Politzer}. At the relatively low
energies, for the electromagnetic field, obviously $C_1=0$. However,
above the  scale of electroweak phase transition, the electromagnetic
fields mix with other vector bosons and become part of the
asymptotic freedom scheme. Thus, depending on the energy scale
both signs are, in principle, possible. We assume that the one-loop
effects are dominating and ignore the higher loop effects, except
for the discussion in the last sections.

 The trace anomaly in the radiation sector has the form
 (see, e.g., \cite{OUP} for the details)
 \beq
\langle T_\mu^\mu \rangle
\,=\, -\,\frac{2}{\sqrt{-g}}\,g_{\alpha\beta}\,
\frac{\de \Ga_r}{\de g_{\alpha\beta}}
\,=\,
- \frac{1}{\sqrt{-{\bar g}}}\,
\frac{\de\,\Ga_r[{\bar g}_{\al\be}\,e^{2\si}]}{\de \si}
\,\bigg|_{{\bar g_{\al\be}}\rightarrow g_{\alpha\beta}, \,
\si\rightarrow 0}
\,=\, \frac{\be}{4}\, g^2 F^2\,,
\label{anomaly}
\eeq
where $\,F^2=F_{\mu\nu}F^{\mu\nu}\,$
is the square of the gauge field strength tensor
and $\Ga_r$ is the one-loop
renormalized effective action in the radiation sector. Also, we
introduced the parametrization of the metric
\beq
g_{\al\be} = {\bar g}_{\al\be}\,e^{2\si}= {\bar g}_{\al\be}\,a^2,
\label{gbar}
\eeq
which will prove useful below. In the homogeneous and isotropic
Universe, the unique spacetime coordinate is the conformal time
$\eta$, related to the physical time $t$ by the formula
$dt = a(\eta)d\eta$.
We shall write the next few formulas in a covariant way
and then switch to the flat-space cosmological metric.

Equation~(\ref{anomaly}) can be used to find a solution to the effective
action, and its covariant nonlocal form \cite{rie,frts84} (see also
further developments in \cite{Mottola-08} and \cite{RadiAna}) is
\beq
\Ga_r \,=\,-\, \frac{\,\be g^2}{16\,\,}\iint_{x,y}
\Big(E_4 - \frac23\,\cx R\Big)_x
\,G(x,y)\,F^2(y)\,,
\label{Gar}
\eeq
where $E_4= R_{\mu\nu\al\be}^2 - 4 R_{\al\be}^2 + R^2$ is the
Gauss-Bonnet invariant, $G(x,y)$ is the conformally
covariant Green function of the operator
$\De_4 \,=\, \Box^2 + 2R^{\mu\nu}\na_\mu\na_\nu
- \frac23\,R\Box  + \frac13\, (\na^\mu R)\,\na_\mu$ and
$\int_x = \int d^4x\sqrt{-g(x)}$.
It is possible to formulate the induced action
in the covariant local form \cite{rie}, including with two
auxiliary scalar fields \cite{a}. The last is the most useful
formulation for many applications, such as the classification of
vacuum states \cite{balsan} or the reaction of gravitational
waves to the presence of higher derivatives \cite{AnJu-wave}.
A qualitatively similar representation, with certain simplifications
\cite{Mottola-08,RadiAna},
should be most useful for the analysis of cosmological
perturbations. We leave this part for future work and,
in the rest of this paper, will restrict the consideration by
the basic elements of the cosmological model, i.e., the
dynamics of the homogeneous and isotropic Universe. In this
case, one can use a much simpler form of induced action,
which is equivalent to (\ref{Gar}) for this special metric.

In covariant form, the anomaly-induced term mixes radiation
and curvature-dependent terms.
In the cosmological framework, we assume that the fiducial
metric is flat and then (\ref{Gar}), with an additional
Einstein-Hilbert term and cosmological
constant, boils down to the non-covariant local form
\beq
\Ga
\,=\, -\,\frac{1}{16\pi G}\int d^4x\sqrt{-g}\,\big(R+2\La\big)
\,-\, \frac{\,\be g^2}{4\,}  \int d^4x\,\sqrt{-\bar{g}}\,\bar{F}^2\si,
\label{Action}
\eeq
where the bars denote quantities defined
using the fiducial metric,
$\bar{g} = \det \big( \bar{g}_{\mu\nu} \big)$ and
$\bar{F}^2 = \bar{g}^{\mu\al} \bar{g}^{\nu\be}
F_{\mu\nu}F_{\al\be}$.
It is fairly easy to check that the last term in
(\ref{Action}) is a solution of (\ref{anomaly}).

\section{The bounce solution}
\label{sec3}

Let us consider an analytical cosmological solution in the theory
(\ref{Action}) and use it for making general conclusions that go
beyond QED and even beyond the Standard Model.

Taking the variational derivative with respect to $\si(\eta)$ and
changing the variables to the physical time $t$ and
$a(t) = \exp \big\{ \si(t)\big\}$, we arrive at the equation
(note the change of notations compared to \cite{RadiAna})
\beq
&&
\frac{\ddot{a}}{a}
+ \frac{\dot{a}^2}{a^2}
\,=\,
\frac{\mathcal{M}}{2a^4} + \frac{16 \pi}{3M_P^2} \, \rh_\La .
\label{eom_a}
\eeq
In this expression and below, we use the notations
\beq
M_P^2 \, = \, \frac{1}{G},
\qquad
\mathcal{M} \, = \, \frac{2 \pi \be g^2}{3M_P^2}\,{\bar F}^2,
\qquad
\rho_\La \, = \, \frac{\La}{8 \pi G}\,.
\label{MPM}
\eeq
Previously, the bounce with a cosmological constant had been
considered, e.g., in \cite{BoNePa12}.
Our expression (\ref{MPM}) includes the cosmological constant term
and the anomalous part described above. Usually, one can assume that
the cosmological constant is irrelevant at extremely high energies
where the cosmological singularity or a bounce should take place.
On the other hand, at the energies above the scale of the electroweak
phase transition, the cosmological constant is supposed to change
its magnitude by many orders \cite{Weinberg89}. Regardless of the
cosmological term $\rho_\La$ is subleading \cite{BludRud} compared
to the radiation energy density, we take it into account. That is
especially important because classical radiation
does not enter directly Eq.\,(\ref{eom_a}) and, as we shall see in
a moment, shows up only after the first integration.

The first integration, or order reduction, can be done by taking
the Hubble parameter as a function of the conformal factor
$H(a) = \dot{a}/a$. This approach brings the relation
\beq
H^2 \,=\, \frac{C}{a^4}
\,+\, \frac{\mathcal{M}}{a^4}\,\log \frac{a}{a_0}
\,+\, \frac{\La}{3}\,.
\label{Hubble-pre}
\eeq
It is worth noting that, in the classical approach, there are 
both a dynamical equation for the scale factor and a constraint
equation. In case of the terms generated by trace anomaly in both
gravitational and radiation sectors, the constraint equation also
takes place in the form $\langle \na_\mu T^{\mu\nu}\rangle = 0$,
reflecting the general covariance of the anomaly-induced
action, including the nonlocal term (\ref{anomaly}). This part was
elaborated in detail in Ref.~\cite{RadiAna}, so we can skip the
details and just give the equation for the pressure of radiation in
the presence of the anomaly, supplementing the energy density
equivalent to (\ref{Hubble-pre}),
\beq
p_r
\,=\,\frac13\,\Big(
\rho_r \,-\, \frac14\,\frac{|\be|g^2{\bar F}^2}{a^4}\Big)\,.
\label{pressure}
\eeq
The last term on the \text{rhs} represents a quantum correction
to the equation of state for the radiation.
In the present work, we will not use this expression, but it may
be useful for the analysis of cosmic perturbations. On the other
hand, a more solid approach should be based on the local version
of the covariant expression (\ref{Gar}).

Coming back to our consideration, the second term in the 
\text{rhs} of (\ref{Hubble-pre}) vanishes
in the classical limit, and this enables us to identify the integration
constant $C$ with the product $\rho_{r0} a_0^4 M_{P}^{-2}$,
where $\rho_{r0}$ is a radiation
energy density at $a = a_0$. The comparison with our previous
parametrization of the metric (\ref{gbar}) makes us assume that
$\bar{g}_{\mu\nu}$ corresponds to the value of $a_0$.
Consequently, we replace $\bar{F}^2 \to F_0^2a_0^4$ in the
formula for  $\mathcal{M}$ in (\ref{MPM}). After that, the
previous relation (\ref{Hubble-pre}) is cast into the form
\beq
H^2 \,=\, \frac{a_0^4}{a^4}\,\bigg(
\frac{\rho_{r0}}{M_{P}^{2}}
\,+\, {\mathcal{M}}\,\log \frac{a}{a_0}\bigg)
\,+\, \frac{\La}{3}\,.
\label{Hubble}
\eeq

In all these relations, the value $a_0$ corresponds to the size
of the Universe where our approximations apply. That means,
$a_0$ should provide sufficiently high energies to have either
\ \textit{i)} a radiation-dominated regime, when the role of massive
particles (in the form of dust or larger objects) is irrelevant
compared to radiation; or \ \textit{ii)} all matter particles 
at such high temperatures that their masses are negligible.

Thus, the questions to address are as follows:
\textit{a)} whether Eq.~(\ref{Hubble}) admits
an analytic solution corresponding to a bounce, and
\textit{b)} if the required difference in size between
$a_0$  and the value $a_m$ corresponding to a bounce,
takes us to the trans-Planckian energies.
The successful bounce model should answer negatively to the last
question, as otherwise, we cannot justify ignoring the quantum
gravity effects. Here we consider part \textit{a)} and leave the
more complicated question \textit{b)} to the next section.

\subsection{Analytic solution for a bounce}

As we know \cite{BludRud}, for a sufficiently small $a_m \ll a_0$,
the cosmological term is small compared to other terms on the
\text{rhs} of  (\ref{Hubble}). Thus, we can explore the bounce
solution for $\La=0$ and then include a non-zero $\La$-term, treating
it as a small perturbation. In this way, using (\ref{MPM}), we arrive
at the condition of $H(a_m)=0$ in the form
\beq
\rho_{r0}
\,=\, -\, {\mathcal{M}}M_P^2\,\log \frac{a_m}{a_0}
\,=\, \frac{2\pi}{3}\, \be g^2\,F_0^2 \,\log \frac{a_0}{a_m}\,.
\label{Hubble_zero}
\eeq
Since we suppose that the Universe is initially contracting,
$a_0 > a_m$ and hence the necessary condition of the bounce
is that $ \be F_0^2 > 0$.

As the first example, consider the simplest case when the Universe
is very hot and its contents can be described by the energy density
of radiation $\rho_r$. On the other hand, the space is conducting
owing to the presence of a hot gas of charged particles. For the
sake of simplicity, we assume that, in the initial point of the
relevant phase of the contracting Universe, most of  $\rho_{r0}$
consists of the electromagnetic radiation \cite{Dolgov93}.
Then we have
\beq
\rho_r \approx \frac{\vec{H}^2 + \vec{E}^2}{2}
\qquad
\mbox{and}
\qquad
F^2 \approx \frac{\vec{H}^2 - \vec{E}^2}{2}\,.
\label{rhoF2}
\eeq
Owing to the conducting media, $\vec{E}^2 \approx 0$ 
we arrive at the estimate $\rho_{r0}\approx F_0^2$. Thus, we
arrive at the solution for $a_m$ in the form
\beq
a_m
\,=\, a_m (\La = 0)\,=\, a_0 \exp\Big\{
- \,\frac{3}{2\pi \be g^2}\Big\}\, .
\label{am}
\eeq

Another possibility is to use relation (\ref{Hubble}) with
$\La=0$ and get the general solution
\beq
\label{sol_eq}
&&
t-t_0\,=\,
\pm\,\sqrt{\frac{\pi}{2\mathcal{M}}}\,e^{-2C/\mathcal{M}}
\left[\textrm{erfi}\left(\sqrt{2\log a+2C/\mathcal{M}}
\right)-\textrm{erfi}\left(\sqrt{2C/\mathcal{M}}\right)
\right],
\quad
\eeq
where
$\textrm{erfi}\left(x\right)=-i\,\textrm{erf} \left(ix\right)$
is the imaginary error function. Treating  $\mathcal{M}$ as a
small perturbation, we can use the asymptotic expansion
\beq
&&
\textrm{erfi}\left(x\right)\Big|_{x\rightarrow\infty}\simeq
-\,i+
\frac{e^{x^{2}}}{\sqrt{\pi}\,x}
+O\left(x^{-1}\right),
\eeq
and derive the following approximate solution:
\beq
&&
t\simeq
\pm\,\frac{1}{2\sqrt{C}}\left[
a^{2}\left(1-\frac{\mathcal{M}}{2C}\log a\right)
-1\right],
\;\;\;\;\;\;\;\;\;t_{0}=0.
\eeq
In the limit $t\rightarrow0$, we verify that \ $a(t)\rightarrow1$,
which is consistent with the numerical solutions, as we will see
in the next subsection. Additionally, taking the limits
$t\rightarrow\pm\infty$, we find $a(t)\rightarrow+\infty$.
Let us note that this scheme is opposite to what is required for
the bounce since, in the last case,  $\mathcal{M}$ cannot be
regarded as small.

Taking the cosmological constant term as a small perturbation
in Eq.~(\ref{Hubble}) is a technically simple exercise, and we give
only the final result:
\beq
a_m(\La)
\,=\, a_m
\Big(1 \,-\, \frac{4\pi\,\rho_\La}{\be g^2 F_0^2}\,
\frac{a_m^4}{a_0^4}\,\Big)\, .
\label{amCC}
\eeq
Typically, this formula describes
a small correction to the basic solution (\ref{am}).

\subsection{Plots corresponding to the bounce}

\begin{figure}[ht!]
\centering
\includegraphics[width=0.472\textwidth]{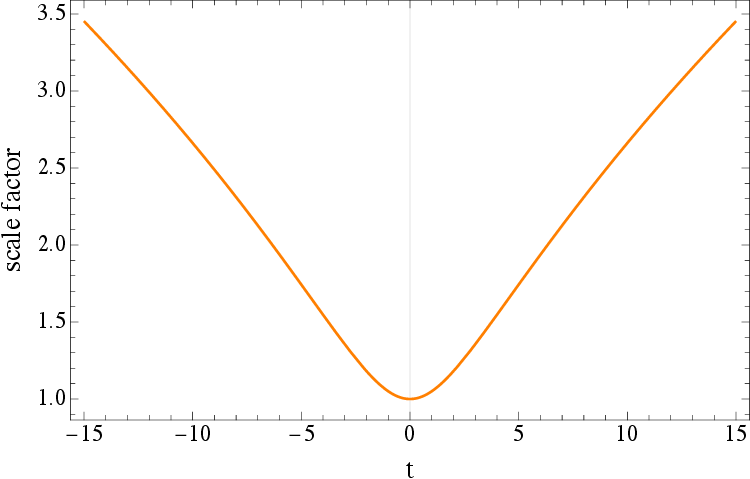}
\quad
\includegraphics[width=0.491\textwidth]{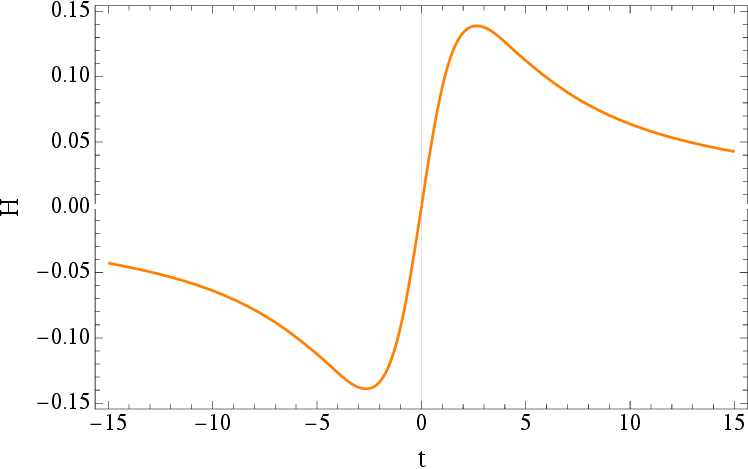}
\vspace{-8.5mm}
\caption{\small
Numerical solution for the scale factor $a(t)$ in the presence of the 
anomalous radiation term.
We assumed the value $\be g^2 F_0^2 = 0.1$  in the Planck units
and the initial conditions $a(0) = 1$ and $\dot{a}(0)=-10^{-3}H_0$.
The left plot shows $a(t)$ in the range $-15 \leq t \leq 15$.
The right plot shows the Hubble parameter $H(t)$.
}
\label{Fig2}
\end{figure}

Let us first illustrate the analytic solution presented above by a
few plots obtained by the numerical solution of
Eq.~\eqref{eom_a} with $\rho_\La = 0$ using Mathematica
\cite{Wolfram}.
Imposing the initial conditions corresponding to contraction,
one arrives at the bounce
type plots of $a(t)$, with a smooth transition between the
contracting and expanding phases. These plots are shown in
Figs.~\ref{Fig2}. The last curve clearly shows that the Hubble
parameter $H$ evolves smoothly through the bounce region.

\begin{figure}[ht]
\centering
\includegraphics[width=0.488\textwidth]{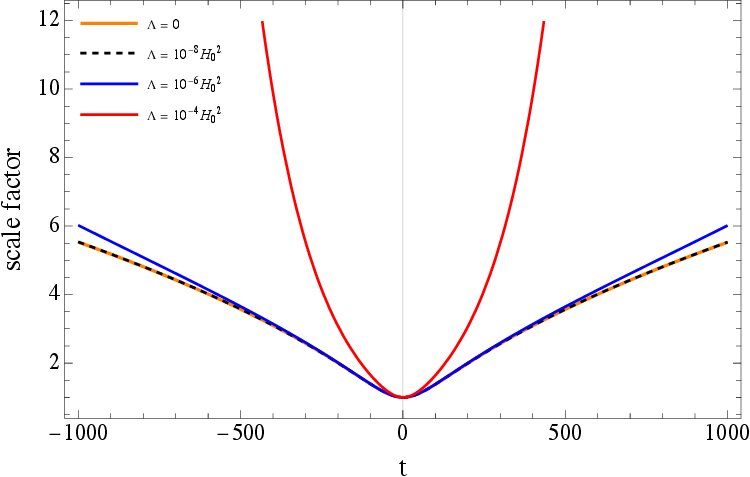}
\quad
\includegraphics[width=0.475\textwidth]{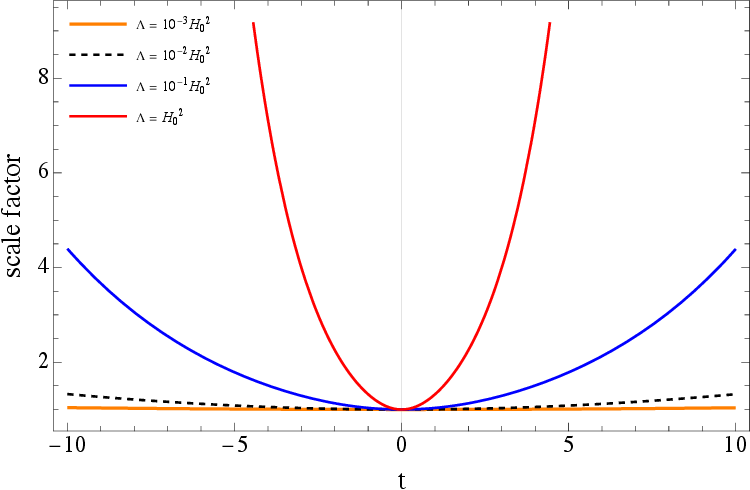}
\vspace{-8.5mm}
\caption{\small
Numerical solution for the scale factor $a(t)$ in the presence of the 
anomalous radiation term and positive cosmological constant.
The values used for getting the numerical solutions are indicated
on the plots.}
\label{Fig3}
\end{figure}

The last point concerning the solutions without the cosmological
constant is that the general shape of the bouncing solutions does not
depend on the values of parameters and on the details of initial
conditions. Thus, the analytic results from the previous subsection
are perfectly well confirmed and there are no issues with the
stability in this model of the bounce.
When the cosmological constant term is positive, the numerical
analysis shows that the bounce-type solutions remain. However, with
the growth of the magnitude of $\La$, the plots become narrow. The
plots obtained with different values of $\La$ and ranges of $t$, are
presented in Fig.~\ref{Fig3}.
We adopt Planck units for time $t$ in all the plots.

All the mentioned features concern only the positive cosmological
constant. Let me mention that, in the case with the negative
cosmological constant, there is a non-singular cyclic behavior of
$a(t)$. The difference with the known cyclic models  (see, e.g.,
\cite{Steinhardt02}, also \cite{Steinhardt19,BoJu20}, and references
therein) is that, in the present case the frequency of the oscillations
is very high. Since we do not have physical interpretation of this
type of solution, it will not be discussed in detail here.

\section{Quantitative estimates and analysis}
\label{sec4}

The consideration of the physical significance of the bounce solution
starts with the note that, in the expanding or contracting Universe,
the typical energy of a photon, or the temperature of the background
radiation, is inverse to the scale factor, i.e.,
\beq
\frac{a}{a_0}\,=\,\frac{T_0}{T}\,.
\label{Ta}
\eeq
Thus, the solution (\ref{am}) implies the following estimate for
the energy in the bounce point:
\beq
T_m
\,=\, \, T_0 \exp\Big\{
\frac{3}{2\pi \be g^2}\Big\}\, .
\label{amT}
\eeq
Taking the values corresponding to QED, the coupling satisfies
$\al = g ^2/(4\pi) \approx 1/137$, which provides a very pessimistic
estimate of $\,T_m \gg 10^{100} GeV$. This means, the bounce
solution occurs at the energies in the very deep trans-Planckian
regime, i.e., far beyond the framework of the approximation we use.
The conclusion is that the consideration based on QED does not
form a sound basis for the semiclassical bounce model without
additional assumptions. Is it possible to get a better estimate in
more general theories?

The expression (\ref{amT}) is quite sensible to the magnitude of the
product $\be g^2$, owing to the exponential dependence. It is clear
that the numerical estimate for the bounce may be improved in two
ways, namely by increasing the value of $g$ and increasing the beta
function, according to the general formula (\ref{ym19}) and beyond
this formula. According to interpretation \textit{ii)} in
Sec.~\ref{sec3}, we can assume that the temperature $T_0$ is of the
order of grand unification scale $M_X$ or slightly lower, such that
all matter particles have high kinetic energies and their masses are
negligible. Then the definition (\ref{MPM}) should be modified.
Indeed, it is not sufficient to replace
$\bar{F}^2 \to \bar{\mathcal L}$,
where the last symbol indicates the covariant Lagrangian of the whole
theory, including fermions, scalars and vectors, at the point $a_0$.
The reason is that the term $\rho_{r0}$ in the main equation
(\ref{Hubble}), should be interpreted as the energy density of the
whole contents of the Universe at the corresponding high energy
scale. The product $\be g^2 F^2$ should be replaced by the sum
of the terms corresponding to different
parts of the Lagrangian.  Then the expression (\ref{amT}) should
be replaced by
\beq
T_m
\,=\, \, T_0 \exp\Big\{
\frac{3\,\sum_k\bar{\rho}_k}{2\pi \,\sum_k \be_k g_k^2
\bar{\mathcal L}_k}\Big\}\, ,
\label{amGUT}
\eeq
where index $k$ runs over all fields in the Lagrangian. This
expression is model dependent and its evaluation is beyond the
scope of the present work. Let us, anyway, list the requirements
for the acceptable bounce in this framework.

1. To have a sufficiently small ratio in the exponential in
(\ref{amGUT}), at least some of the coupling constants should be
large. That means, a phenomenologically successful bounce without
modification of gravity or a special scalar field requires that at  least
part of the couplings are strong and, consequently, the account of
non-perturbative effects in the corresponding QFT.

2. The sign of the denominator in the exponential in
(\ref{amGUT}) should be positive as otherwise equation
(\ref{Hubble}) would not have bounce solutions.

3. The magnitude of the ratio in the exponential in (\ref{amGUT})
should be such that $T_m$ belongs to the interval between the masses
of at least some of the quantum particles and the Planck scale, where
we assume modifications of the action of gravity and, probably,
quantum gravitational effects.

4. To provide a correspondence with the observational data concerning
inflations, it is important that the bounce region starts and ends with
a very high $|H_0|$, e.g., in the interval $10^{11} - 10^{13}\,GeV$.
Without modifying the gravitational action, this means that the
initial point $a_0$ corresponds to the temperature (typical energy)
$T_0  \sim \big[H_0^2 M_P^2\big]^{1/4}$.  On the other hand,
the simplest description of inflation is the Starobinsky model
\cite{star}, that
corresponds to adding the $R^2$-term with the coefficient about
$5 \times 10^8$ \cite{star83}. We plan to explore this extension
of the model described above in the future work \cite{SWS}, but
assuming that this extra term does not have a dramatic effect on
the value of $T_0$, we arrive at the narrow interval of Hubble
parameters $-|H_0| < H < |H_0|$ and the temperatures
$T_0 < T < M_P$.

The last observation concerns the first of the listed points. In the
case of strong coupling, the one-loop approximation which we
used here is not appropriate. The required modifications  do not
reduce to the change of the beta functions and the corresponding
modifications in the anomaly. The point is that the first order in
$\si$ in Eq.~(\ref{Action}) and in the similar extended formulas
related to (\ref{amGUT}) reflect only the violation of local
conformal symmetry corresponding to the first logarithms, such
as terms proportional to $\,L = \log \big(\cx/\mu^2\big)$ in the UV
form factors (see, e.g., \cite{OUP} for detailed explanation).

Let us use this information as a hint to what may happen at
higher loops. At the second loop, there is certainly the $\,L^2$-type
addition in the form factor of the $F_{\mu\nu}F^{\mu\nu}$-term
in the electromagnetic sector; in the third loop there will be the 
$\,L^3$-type addition, etc. Let us stress that these extra logarithms
are companions of the leading divergences of the theory before
the renormalization is applied. Thus, since the underlying theory
is renormalizable, the structure of the terms in the action remains
the same, and the changes concern only the form factors. As a
result, the leading-log terms in the non-perturbative regime will
give the complication in the action (\ref{Action}),
\beq
\Ga_{np}
\,=\, -\,\frac{1}{16\pi G}\int d^4x\sqrt{-g}\,\big(R+2\La\big)
\,-\, \frac{1}{4}  \int d^4x\,\sqrt{- \bar{g}}\,\bar{F}^2
\si B(g^2 \si),
\label{Action-np}
\eeq
where $B(x)$ is some unknown function corresponding to
the summation of the perturbative series.\footnote{Assuming
that this series is convergent, in some sense.} Let us note that
the leading logarithmic terms always enter with the coefficient
$g^2$, and the same is true for the powers of $\si$, such that the
argument of $B$ should be the product $g^2 \si$. It is clear that
this modification may change the solution such as (\ref{am}),
including it may wash out the bounce, or modify the shape
of the $a(t)$ dependencies, etc. The only thing we can say
at this point is that the bounce of the described type, completely
based on particle physics and without additional inputs, is
possible. On the other hand, its detailed investigation requires
better knowledge of many issues, such as UV completion
of the Standard Model and summing up the leading logs in
the UV regime.

\section{Conclusions and discussions}
\label{sec5}

We have found an analytic solution describing the cosmological bounce
without modifying the action of gravity, introducing a scalar field,
or accounting for the vacuum quantum effects. The bounce occurs
owing to the equilibrium between the classical radiation term and the
quantum correction in the radiation-gravitational sector (\ref{Gar}).
The form of these loop contributions is well known and does not
require anything besides the well-established results of quantum field
theory. If comparing with the previously known models with bounce,
the anomaly-induced correction to radiation plays the role
of the phantom scalar \cite{BoNePa02}.

The numerical estimates show that, in the minimalist QED
framework, the bounce occurs at absurdly high energies,
making the aforementioned analytic
solution physically senseless. On the other hand, this estimate
is exponentially dependent on the value of the strongest coupling
constant of the theory. The physically
acceptable bounce is possible, but this imposes strong restrictions
on the underlying particle physics model beyond the Standard
Model. In particular, there should be a UV regime with a strong
coupling, similar to what is required for the fully QFT-based
stable version of Starobinsky inflation \cite{StabInstab}.

The mentioned conditions do not look completely impossible to
satisfy, but, at the present state of the art, it is not feasible to state
that this kind of bounce is a realistic scenario to avoid singularity.
Anyway, we can conclude that, in principle, the semiclassical
effects in the radiation sector at the GUT scale may provide the
singularity avoidance without additional \textit{ad hoc} assumptions.

Another open question is the stability of the bounce model under
discussion under the density and metric perturbations. This issue is
typically complicated in all bounce models. The reason is that, in
the vicinity of the bounce, the time derivative $\dot{H}$ is
necessary positive, and this leads to the violation of the null energy
condition (NEC). This feature may lead to instabilities in cosmic
perturbations \cite{Visser1998,Novello2008} (see also
\cite{PatrickNelson01} for an alternative discussion). The analysis
of the cosmological perturbations in a cosmological model is a
necessary element of its development, and this is especially true for
models with a bounce \cite{Peter2015}. Only the analysis of
perturbations may show whether the given model is viable or
possesses inconsistencies.

In general, additional restrictions on the bounce models come from
the feature that the amplitude of the primordial power spectrum is
usually proportional to the energy scale of the bounce. The power
spectrum is constrained by CMB data, which can provide more stringent
constraints on the bounce energy scale. This part was extensively
discussed, e.g., in
\cite{Brandenberger-2012,Peter2015,Brandenberger-2016} and more
recently in \cite{Cai,Paola-Ruth-Nelson}. The general situation is
such that the mentioned constraints can be obtained only on the basis
of the given cosmological model, as they are sensible to the structure
of cosmic perturbations.\footnote{The authors
gratefully acknowledge discussions with Patrick Peter, Sandro
Vitenti, and Alex Vikman on this subject.} Thus, we leave the
investigation of this issue to the next work with the analysis
of perturbations.

In the existing literature, there are strong indications of that the
violation of NEC by quantum corrections may not lead to the
inconsistencies \cite{Ford2003} and that the same is true in the
theories with scalar fields \cite{Rubakov14,IjjStein16}. Both
arguments apply to our case. It is important to note that the
perturbations should be analyzed not on the basis of the simplest
non-covariant form of induced action (\ref{Action}), but using
the covariant form (\ref{Gar}), in the local representation. In this
case, the theory always includes two auxiliary scalar fields
\cite{a,balsan,AnJu-wave} and, therefore, there are chances to arrive
at the consistent model of bounce, including the perturbations free
of pathologies, according to the criterion of \cite{Peter2015}.
We hope to come back to the detailed consideration of this issue
and, as a first step, construct a new simplified formulation of the
induced action with auxiliary fields, in a close future.

\section*{Acknowledgments}

The authors are very grateful to Patrick Peter for his interest in this
work and very useful explanations concerning the role of cosmic
perturbations in bounce models. We also appreciate the help of
Sandro Vitenti and Alex Vikman in explaining to us some details
about CMB restrictions on the bounce models.
W.C.S. thanks CAPES for supporting his Ph.D. project.
The work of I.Sh. is partially supported by Conselho Nacional de
Desenvolvimento Cient\'{i}fico e Tecnol\'{o}gico - CNPq under 
Grant No. 303635/2018-5.





\begin{thebibliography}{99}

\renewcommand{\baselinestretch}{1.1}
\small

\bibitem{Coles} P. Coles and F. Lucchin,
\textit{Cosmology: The Origin and Evolution of Cosmic Structure}
(Wiley-VCH, Second edition, 2002).

\bibitem{Tolman31} R. C. Tolman,
{\it On the problem of the entropy of the Universe as a whole,}
Phys. Rev. {\bf 37} (1931) 1639.
	
\bibitem{Bounce70} T. Ruzmaikina and A.A. Ruzmaikin,
{\it Quadratic corrections to the Lagrangian density of the
gravitational field and the singularity,}
JETP {\bf 30} (1970) 372.

\bibitem{StarBo78} A.A. Starobinsky,
{\it On a singular isotropic cosmological model,}
Sov. Astron. Lett. {\bf 4} (1978) 82.

\bibitem{Novello2008} M.~Novello and S.E.P.~Bergliaffa,
{\it Bouncing cosmologies,}
Phys. Rept. \textbf{463} (2008) 127, 
arXiv:0802.1634.

\bibitem{Peter2015} D.~Battefeld and P.~Peter,
{\it A critical review of classical bouncing cosmologies,}
Phys. Rept. \textbf{571} (2015) 1, 
arXiv:1406.2790.

\bibitem{Biswas2012} T.~Biswas, A.S.~Koshelev,
A.~Mazumdar and S.Y.~Vernov,
{\it Stable bounce and inflation in non-local higher
derivative cosmology,}
JCAP \textbf{08} (2012) 024,
arXiv:1206.6374.

\bibitem{Koshelev2013} A.S.~Koshelev,
{\it Stable analytic bounce in non-local Einstein-Gauss-Bonnet cosmology,}
Class. Quant. Grav. \textbf{30} (2013) 155001,
arXiv:1302.2140.

\bibitem{AnJu} J.C. Fabris, A.M. Pelinson and I.L. Shapiro,
{\it Anomaly-induced effective  action for gravity and inflation,}
Grav. Cosmol. \textbf{6} (2000) 59, gr-qc/9810032;
Nucl. Phys. B Proc. Suppl. {\bf 95} (2001) 78,
gr-qc/9810032.

\bibitem{AnoBo21} W.C. e Silva and I.L.~Shapiro,
{\it Bounce and stability in the early cosmology with
anomaly-induced corrections,}
Symmetry {\bf13} (2021) 50,
arXiv:2012.10554.

\bibitem{KaroubyBo} J. Karouby and R. Brandenberger,
{\it A radiation bounce from the Lee-Wick construction?,}
Phys. Rev. \textbf{D82} (2010) 063532, 
arXiv:1004.4947;
\\
J. Karouby, T. Qiu and R. Brandenberger,
{\it On the instability of the Lee-Wick bounce,}
Phys. Rev. \textbf{D84} (2011) 043505, 
arXiv:1104.3193.

\bibitem{AC} T.~Appelquist and J.~Carazzone,
{\it Infrared singularities and massive fields,}
Phys. Rev.  {\bf D11} (1975) 2856.

\bibitem{fhh} M.V.~Fischetti, J.B.~Hartle, and B.L.~Hu,
{ \it Quantum effects in the early universe. I. Influence of
trace anomalies on homogeneous, isotropic, classical geometries,}
Phys. Rev. {\bf D20} (1979) 1757.

\bibitem{star} A.A. Starobinski,
{ \it A new type of isotropic cosmological models without
singularity,}
Phys. Lett. {\bf B91} (1980) 99.

\bibitem{Dolgov93} A.D.~Dolgov,
\textit{Breaking of conformal invariance and electromagnetic field
generation in the Universe,}
Phys. Rev.  {\bf D48} (1993) 2499.

\bibitem{OUP} I.L.~Buchbinder and I.L.~Shapiro,
\textit{Introduction to quantum field theory with applications
to quantum gravity,}
(Oxford University Press, 2021).

\bibitem{GrossWilczek} D.J. Gross and F. Wilczek,
{\it Ultraviolet behavior of non-Abelian gauge theories,}
Phys. Rev. Lett. {\bf 30} (1973) 1343.

\bibitem{Politzer} H.D. Politzer,
{\it Reliable perturbative results for strong interactions,}
Phys. Rev. Lett. {\bf 30} (1973) 1346.

\bibitem{rie} R.J.~Riegert,
{ \it A non-local action for the trace anomaly,}
Phys. Lett. {\bf B134} (1984) 56. 

\bibitem{frts84} E.S. Fradkin and A.A. Tseytlin,
{\it Conformal anomaly in Weyl theory and anomaly free
superconformal theories,}
Phys. Lett. {\bf B134} (1984) 187.

\bibitem{Mottola-08} M.~Giannotti and E.~Mottola,
\textit{The trace anomaly and massless scalar degrees of
freedom in gravity,}
Phys. Rev. \textbf{D79} (2009) 045014,
arXiv:0812.0351.

\bibitem{RadiAna} A.M. Pelinson and I.L. Shapiro,
{\it On the scaling rules for the anomaly-induced effective
action of metric and electromagnetic field,}
Phys. Lett. {\bf B694} (2011) 467,
arXiv:1005.1313.

\bibitem{a} I.L. Shapiro and A.G. Jacksenaev,
{ \it Gauge dependence in higher derivative quantum
gravity and the conformal anomaly problem,}
Phys. Lett. {\bf B324} (1994) 286.

\bibitem{balsan} R. Balbinot, A. Fabbri and I.L. Shapiro,
{\it Anomaly induced effective actions and Hawking radiation,}
Phys. Rev. Lett. {\bf 83} (1999) 1494, hep-th/9904074;
{\it Vacuum polarization in Schwarzschild space-time by
 anomaly induced effective actions,}
Nucl. Phys. {\bf B559} (1999) 301, hep-th/9904162.

\bibitem{AnJu-wave} J.C. Fabris, A.M. Pelinson and I.L. Shapiro,
{\it On the gravitational waves on the background
of anomaly-induced inflation,}
Nucl. Phys. {\bf B597} (2001) 539, 
hep-ph/0208184.

\bibitem{BoNePa12} R. Maier, S. Pereira, N. Pinto-Neto,
and B.B. Siffert,
\textit{Bouncing models with a cosmological constant,}
Phys. Rev. {\bf D85} (2012) 023508,
arXiv:1111.0946.

\bibitem{Weinberg89}  S. Weinberg,
{\it The cosmological constant problem,}
Rev. Mod. Phys. \textbf{61} (1989) 1.

\bibitem{BludRud} S.A.~Bludman and M.A.~Ruderman,
{\it Induced cosmological constant expected above the phase
transition restoring the broken symmetry,}
Phys. Rev. Lett. {\bf 38} (1977) 255.

\bibitem{Wolfram} Wolfram Research, Inc.,
\textit{Mathematica,} (Version 12.0, Champaign, IL, 2019).

\bibitem{Steinhardt02} P.J. Steinhardt and N. Turok,
\textit{Cosmic evolution in a cyclic universe,}
Phys. Rev. {\bf D65} (2002) 126003,
hep-th/0111098.

\bibitem{Steinhardt19} A. Ijjas and P.J. Steinhardt,
\textit{A new kind of cyclic universe,}
Phys. Lett. {\bf B795} (2019) 666,
arXiv:1904.08022.

\bibitem{BoJu20} I. Torres, J.C. Fabris and O.F. Piattella,
\textit{Bouncing and cyclic quantum primordial universes
and the ordering problem,}
Class. Quantum Grav. {\bf 37} (2020) 105005,
arXiv:1911.01487.

\bibitem{star83} A.A.~Starobinsky,
{\it The perturbation spectrum evolving from a nonsingular initially
de-Sitter cosmology and the microwave background anisotropy,}
Sov. Astron. Lett. {\bf 9} (1983) 302.

\bibitem{SWS} S.W.P. Oliveira, W.C. e Silva and I.L. Shapiro,
{\it Work in progress.}

\bibitem{BoNePa02} P. Peter and N. Pinto-Neto,
\textit{Primordial perturbations in a nonsingular bouncing
universe model,}
Phys. Rev. {\bf D66} (2002) 063509.
hep-th/0203013.

\bibitem{StabInstab}
T.d.P.~Netto, A.M.~Pelinson, I.L.~Shapiro and A.A.~Starobinsky,
{\it From stable to unstable anomaly-induced inflation,}
Eur. Phys. J. {\bf C76} (2016)  544,
arXiv:1509.08882.

\bibitem{Visser1998} C.~Molina-Paris and M.~Visser,
\textit{Minimal conditions for the creation of a
Friedman-Robertson-Walker universe from a 'bounce',}
Phys. Lett. \textbf{B455} (1999) 90, 
gr-qc/9810023.

\bibitem{PatrickNelson01} P. Peter and N. Pinto-Neto,
\textit{Has the Universe always expanded?,}
Phys.Rev. {\bf D65} (2001) 023513,
gr-qc/0109038.

\bibitem{Brandenberger-2012} R.~Brandenberger,
\textit{The matter bounce alternative to inflationary cosmology,}
arXiv:1206.4196.

\bibitem{Brandenberger-2016} R.~Brandenberger and P.~Peter,
\textit{Bouncing Cosmologies: Progress and Problems,}
Found. Phys. \textbf{47} (2017) 797, 
arXiv:1603.05834.

\bibitem{Cai}
Y.-F. Cai, F. Duplessis, D.A. Easson and D.-G. Wang,
\textit{Searching for a matter bounce cosmology with low
	redshift observations,}
Phys. Rev. {\bf D93} (2016) 043546,
arXiv:1512.08979.

\bibitem{Paola-Ruth-Nelson}
P.C.M. Delgado, R. Durrer and N. Pinto-Neto,
\textit{The CMB bispectrum from bouncing cosmologies,}
JCAP {\bf 11} (2021) 024,
arXiv:2108.06175.

\bibitem{Ford2003} L.H.~Ford,
\textit{The classical singularity theorems and their
quantum loopholes,}
Int. J. Theor. Phys. \textbf{42} (2003) 1219, 
gr-qc/0301045.

\bibitem{Rubakov14} V.A. Rubakov,
\textit{The null energy condition and its violation,}
Phys. Usp. {\bf 57} (2014) 128,
arXiv:1401.4024.
	
\bibitem{IjjStein16} A. Ijjas and P.J. Steinhardt,
\textit{Classically Stable Nonsingular Cosmological Bounces,}
Phys. Rev. Lett. {\bf 117} (2016) 121304,
arXiv:1606.08880.

\end{thebibliography}
\end{document}